%% file: acl_latex.tex
\pdfoutput=1

\documentclass[11pt]{article}

\usepackage[final]{acl}

\usepackage{times}
\usepackage{latexsym}
\usepackage{enumitem}

\usepackage[T1]{fontenc}

\usepackage[utf8]{inputenc}

\usepackage{microtype}

\usepackage{inconsolata}

\usepackage{graphicx}

\usepackage{tabularx}
\usepackage{xcolor}
\usepackage{booktabs}

\usepackage{listings}
\usepackage{xcolor}

\lstdefinestyle{mystyle}{
    language=Python,
    basicstyle=\ttfamily\small,
    keywordstyle=\color{blue}\bfseries,
    stringstyle=\color{teal},
    commentstyle=\color{gray},
    morestring=[b]",
    showstringspaces=false,
    breaklines=true,
    frame=single
}

\usepackage{multirow}

\title{Transforming Podcast Preview Generation: From Expert Models to LLM-Based Systems}

\author{Winstead Zhu \and Ann Clifton \and Azin Ghazimatin \and Edgar Tanaka \and Ward Ronan \\ Spotify \\ \texttt{\{winsteadx,aclifton,azing,edgart,edwardr\}@spotify.com}}

\begin{document}
\maketitle
\begin{abstract}
Discovering and evaluating long-form talk content such as videos and podcasts poses a significant challenge for users, as it requires a considerable time investment. Previews offer a practical solution by providing concise snippets that showcase key moments of the content, enabling users to make more informed and confident choices. We propose an LLM-based approach for generating podcast episode previews and deploy the solution at scale, serving hundreds of thousands of podcast previews in a real-world application. Comprehensive offline evaluations and online A/B testing demonstrate that LLM-generated previews consistently outperform a strong baseline built on top of various ML expert models, showcasing a significant reduction in the need for meticulous feature engineering. The offline results indicate notable enhancements in understandability, contextual clarity, and interest level, and the online A/B test shows a 4.6\% increase in user engagement with preview content, along with a 5x boost in processing efficiency, offering a more streamlined and performant solution compared to the strong baseline of feature-engineered expert models. 

\end{abstract}

\input{sections/introduction}

\input{sections/related-work}

\input{sections/system-design}

\input{sections/experiments}

\input{sections/results}

\input{sections/conclusion}

\input{sections/ethics-statement}

\bibliography{custom}

\appendix

\input{sections/appendix_timestamped_transcript}

\end{document}

%% file: sections/introduction.tex
\section{Introduction}
Podcasts, videos, and other long-form talk content, have become flourishing media, offering diverse content that caters to a wide range of audiences. Discovering new content, however, remains challenging, as the long-form nature of episodes demands significant time investment to assess their relevance~\cite{jones2021current}. Previews, which are short and representative segments of an episode, provide a solution by capturing engaging, self-contained moments that are easy to understand without additional context~\cite{barua2025lotus}.

Generating effective previews from episodes that can exceed an hour is a challenging task and requires robust content understanding. For example, to locate self-contained segments, previous work suggests using segmentation methods to detect topic transitions~\cite{lukasik2020text, liu2022end, retkowski2024text, podtile}. These methods, however, may miss many interesting moments depending on the granularity of segmentation. Furthermore, they typically fail to distinguish between segments containing commercial content such as ads or self-promotions which do not represent the whole content.

Traditional 
preview extraction
approaches often rely on sophisticated feature engineering to derive aggregations of expert models such as sentiment analysis, topic modeling, speech classification, and ad detection, which can be resource-intensive and time-consuming~\cite{rui2000automatically, dabholkar2016automatic}. 
In the meantime, the advent of large language models (LLMs) have transformed the landscape of content understanding and curation~\cite{salemi2023lamp, kirstein2024tell, manatkar2024quis}. 

In this paper, we propose leveraging large language models (LLMs) to extract short, compelling and self-contained episode previews. Using only text-based inputs, including episode metadata (title and description) and transcript, we employ few-shot learning with curated examples of high-quality previews to guide LLMs in identifying the characteristics of an effective preview. To extract the preview segment, we prompt the LLMs to provide structured outputs, specifying start and end timestamps to define precise boundaries.


Our contributions are threefold. First, we successfully integrate an LLM into a large-scale, real-world application for podcast preview extraction. 
Secondly, we propose to use sentence indexing and sentencization to effectively analyze and index lengthy podcast transcripts and accurately retrieve LLM-selected previews.
Finally, we demonstrate significant performance improvements over strong baseline expert models through offline human evaluations and online A/B testing, while achieving a 5x improvement in processing efficiency. By showcasing the successful productionization of this novel application of LLMs, we aim to advance the discourse on leveraging language models for complex content processing tasks, highlighting their potential to simplify workflows and enhance performance in real-world applications.


%% file: sections/related-work.tex
\section{Previous Work}

In this section, we highlight previous related work on highlight extraction, document summarization, and podcast preview extraction.

\subsection{Highlight Extraction}
Generating previews for podcast episodes closely parallels the tasks of highlight extraction~\cite{sun2014ranking, badamdorj2021joint, liu2022umt, jie2024unsupervised}. Highlights are typically annotated by human experts~\cite{collins2017supervised, lei2021detecting} or inferred through weakly supervised signals, such as identifying frequently edited segments in videos~\cite{sun2014ranking}. 

Given the domain dependency of labeled data and the cost of gathering them for long-form content, unsupervised approaches for highlight detection have also been explored. These include leveraging aesthetic features~\cite{song2016click} (e.g., selecting visually pleasing thumbnails), detecting recurring audio-visual patterns~\cite{islam2024unsupervised} (e.g., cheering or clapping in sports videos), or employing methods like k-means clustering~\cite{song2016click} or graph-based techniques~\cite{ErkanR04} to identify representative parts of the text or video. While podcast preview generation is similar to highlight extraction, it introduces an additional challenge: the previews must serve as standalone content, providing listeners with a self-contained piece of the content that can be understood on its own.

\subsection{Document Summarization}
Previous studies on document summarization highlight LLMs' power to identify and retrieve \textit{key information} from lengthy documents using both extractive summarization~\cite{zhang2023extractive, chhikara2024lamsum} and abstractive summarization~\cite{TanakaMultilingualPodSumm, chang2024booookscore} approaches. Building on this foundation, we utilize LLMs to extract previews, focusing specifically on extracting contiguous segments of text. However, effective methods are essential for locating and extracting information from \textit{long} texts, and \citealt{podtile} illustrate successful indexing mechanisms for LLM-selected chapters, essential for accurate retrieval from long texts which inspires our work.



\subsection{Traditional vs. LLM-Powered Podcast Preview Extraction}
Previously, preview extraction relied on sophisticated feature-engineered systems, requiring the aggregation of one or more expert models such as sentiment analysis model and emotion recognition model~\cite{Zhu1576969HotSpot, movie_trailer_ibm, automatic_trailer_generation, anomoly_detection}. Our work has been inspired by such traditional methods to focus on human perception-related aspects like sentiment and attention in initial prompts. However, we are able to outperform these traditional approaches by utilizing LLMs, which automate and improve the extraction and retrieval process through implicit prompt iteration for a more nuanced understanding of transcript content.

Overall, by focusing on LLM-powered methodologies, our work advances beyond conventional systems, offering a more streamlined and contextually aware approach for podcast preview extraction and information retrieval.

%% file: sections/system-design.tex
\section{System Design}
In this section, we describe two systems for podcast preview generation: the sophisticated, feature-engineered legacy machine learning (ML) preview extraction system, and the newly developed LLM preview extraction system.

\subsection{Language Filtering}

Currently, both the legacy ML and LLM preview systems have been primarily developed on English podcast data, therefore language filtering is applied in both systems to only process English-language episodes. The legacy ML preview system employs an audio-based in-house model to perform spoken language identification similar to \citealt{zhu23c_interspeech}. The results are then combined with metadata language annotations (which can be noisy) to co-determine the episode language and filter for only English-language episodes. In contrast, the LLM preview system relies solely on existing noisy metadata language annotations for filtering English episodes, without needing extra language detection techniques.

\subsection{Legacy ML Preview System}

\begin{figure*}[t]
  \centering
  \includegraphics[width=\linewidth]{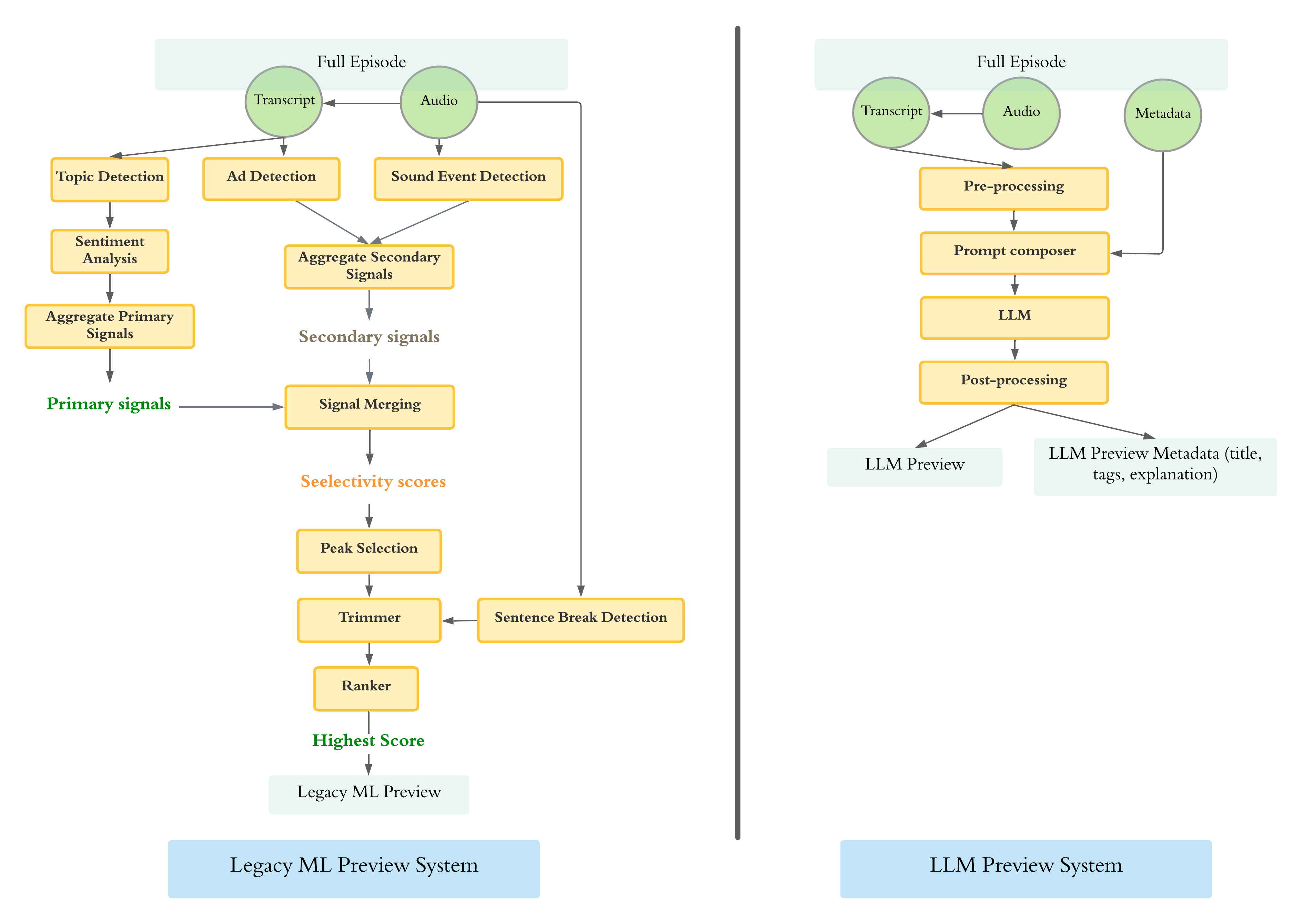}
  \caption {Legacy ML preview system vs. LLM preview system.}
  \label{fig:legacy_ml_system}
\end{figure*}

The legacy ML preview system (Figure \ref{fig:legacy_ml_system} left) is a sophisticated system that utilizes advanced feature engineering and a series of expert models to generate podcast previews. This system involves multiple stages of data processing and signal analysis to select previews. The main components are as follows:

\textbf{Topic Analysis, Sentiment Analysis and Primary Signal Aggregation} The podcast episode transcript is first analyzed using an in-house model to identify key topics, which are then processed by another in-house model to assess the sentiment intensity related to each topic. This creates the so-called \textit{primary signals}, which are aggregated to compute topic trends and identify dominant topics. Speaker boundaries and question-answer segments are also analyzed based on the transcript to ensure smooth transition of topics.

\textbf{Ad Detection and Sound Event Detection:} In parallel to primary signal extraction, the system uses an in-house ad detection model to identify ad content from transcript and an in-house sound event detection model to detect non-speech regions from episode audio. These create the so-called \textit{secondary signals}, which are scaled and aggregated based on predetermined adjustment scores for different types of non-core speech elements such as ads or music.

\textbf{Signal Merging and Peak Selection:} By processing primary and secondary signals, the system derives overall \textit{selectivity scores}. These scores are analyzed to locate peak regions approximately 60 seconds long, from which previews may be extracted.

\textbf{Sentence Break Detection, Trimming and Ranking:} An in-house technique is applied on the episode audio to identify sentence starts and ends, which are considered suitable candidates for preview starts and ends. The detected sentence breaks are then combined with the \textit{selectivity scores} and fed into an in-house trimmer model to adjust the start and end of each preview candidate to improve coherence and context while adhering to the 1-minute duration requirement. Lastly, all preview candidates are ranked by an in-house ranking model to assign a score for each preview candidate. The candidate with the highest ranking score is used as the final preview for the episode.

This sophisticated legacy ML preview system showcases the extensive feature engineering and model integration necessary to produce high-quality podcast previews.

\subsection{LLM Preview System}
While we use a variety of models at Spotify, for this particular use case we use Gemini 1.5 Pro\footnote{Gemini 1.5 Pro: \url{https://ai.google.dev/gemini-api/docs/models/gemini\#gemini-1.5-pro}} in the LLM preview system (Figure \ref{fig:legacy_ml_system} right) to generate podcast previews. Below are the key steps of the system:

\textbf{Pre-processing:} We first sentencize the podcast transcript using simple heuristics such as punctuation markers and annotate each sentence with start and end timestamps in seconds. This step is crucial for enabling the LLM to accurately identify and retrieve the desired preview offset. These timestamped sentences, along with episode metadata such as title and description, form part of the input prompt to the LLM. Appendix \ref{sec:timestamped_transcript} provides a mock example of a pre-processed, sentencized, and timestamped transcript.

\textbf{Preview Offset Selection and Preview Metadata Generation:} We then apply the LLM to perform preview selection and metadata generation. The LLM prompt for preview offset selection incorporates three key elements: 
\begin{enumerate}[topsep=0pt,itemsep=0ex,partopsep=0ex,parsep=0ex]
\item \textbf{Structured reasoning process:} The prompt guides the LLM through a step-by-step structured reasoning process. It begins by examining the episode's title, description, and transcript to identify the main topic. The LLM then evaluates preview segments for relevance and engagement. As part of this structured reasoning, the LLM also generates preview metadata, including a concise explanation of the preview's engagement value and a list of topic tags. This structured reasoning approach not only enhances preview relevancy but also makes LLM's decision-making process more transparent and informed. 
\item \textbf{Preview requirements:} A list of requirements are included in the prompt to ensure that the preview begins with an engaging introduction, progresses logically from foundational concepts to detailed insights, excludes ad content, and starts and concludes with complete thoughts, while aligning with the episode's central theme, evoking emotional resonance, and providing valuable insights. The preview is also required to be approximately one minute long, maximizing audience engagement during normal attention span (\citealt{attention_span}).
\item \textbf{Few-shot learning:} A series of manually curated preview examples are included in the prompt to guide the LLM in learning what constitutes a good preview through few-shot learning \cite{brown2020languagemodelsfewshotlearners}.
\end{enumerate}

\textbf{Prompt Iteration Process:} We manually optimized the prompt to achieve strong alignment with human judgment on a small evaluation dataset of episodes from diverse categories, ensuring broad applicability and generalization across different content types. During the prompt iteration process, feedback was gathered directly from the product and design teams as it was challenging to use automated prompt engineering to replace human input in this case, particularly product and design experts. The process involved iteratively adding and deleting preview requirements and few-shot examples, with human experts re-evaluating the prompt on the small evaluation dataset after each major change to ensure improvement and alignment with human judgment. This manual prompt iteration process, despite not being automated and requiring human oversight, effectively replaced the feature engineering process of the legacy ML system, as it allows for easier incorporation of human feedback, significantly enhancing flexibility and speed when adapting to new preview requirements.

\textbf{Post-processing:} To maintain a concise and coherent preview duration, the preview selected by the LLM is trimmed to the last complete sentence that starts within one minute. While both the LLM-selected preview and the legacy ML system's preview don't always guarantee a one-minute length, the average LLM preview is around 62 seconds long, and the legacy ML preview is around 56 seconds. This means their average durations are not too far apart. Additionally, the need for post-processing trimming is primarily driven by product requirements.

\subsection{Comparison of LLM and Legacy ML Preview Systems}

In comparison, the LLM system offers several advantages over the legacy ML system:

\begin{enumerate}[topsep=0pt,itemsep=0ex,partopsep=0ex,parsep=0ex]
\item \textbf{Streamlined Iterations and Adaptations}: Prompt engineering with LLMs is significantly faster and more streamlined than manual feature engineering and expert model aggregation, as LLMs allow for iterative refinement and quick adaptations to changing requirements.
\item \textbf{Lower Maintenance Complexity}: The legacy ML system involves multiple models and dependencies, making maintenance more complex. In contrast, the LLM system utilizes a single LLM framework, reducing complexity and maintenance effort.
\item \textbf{Faster Processing Speed}: The legacy ML system requires processing audio directly, which is inherently slower compared to processing text. In comparison, the LLM system works primarily with text data and benefits from faster processing times. Both systems have been deployed on the Dataflow streaming platform\footnote{Dataflow streaming pipelines: \url{https://cloud.google.com/dataflow/docs/concepts/streaming-pipelines}}: The legacy ML system takes an average of 100 seconds to process an episode, and the LLM system takes an average of less than 20 seconds; even though both systems are already very fast, the LLM system processes episodes faster, resulting in a 5x improvement in processing time and significantly enhancing scalability.
\end{enumerate}

These improvements highlight the LLM system's advantages in terms of simplicity and scalability, making it a more streamlined solution for generating podcast previews. 

%% file: sections/experiments.tex
\section{Experiments}
In this section, we describe two experiments that we have conducted to evaluate the proposed LLM previews against the legacy ML previews: an offline human evaluation and an online A/B test.

\subsection{Offline Human Evaluation}

We recruited around 20 evaluators internally to evaluate LLM previews against legacy ML previews, and we used Label Studio\footnote{Label Studio: \url{https://labelstud.io/}} platform for data collection and human annotation.

\textbf{Evaluation Setup:} Each evaluator was asked to evaluate around 20 episodes (the actual number of episodes per evaluator was determined based on the time they were able to commit). For each episode, the evaluator was provided with episode metadata including episode title, episode description, and show name, as well as a legacy ML preview and an LLM preview, which were randomly shuffled to prevent position bias favoring one variant over the other. The evaluator then listened to both previews with subtitles, and was asked to choose the better one or indicate a tie.

\textbf{Specific Assessment Questions:} After selecting a preferred preview for a given episode, the evaluator was asked to rate both previews based on three specific questions to understand the relative performance of both systems:
\begin{enumerate}[topsep=0pt,itemsep=0ex,partopsep=0ex,parsep=0ex]
\item \textit{Understandability:} Whether the preview helps determine the episode's relevance for the listener.
\item \textit{Contextual Clarity:} Whether the preview, along with metadata, provides sufficient context to grasp what is being discussed.
\item \textit{Interest Level:} Whether the preview highlights an interesting segment of the episode.
\end{enumerate}
In Table \ref{eval_questionnaire} we present the actual questionnaire that we created for each evaluator on the Label Studio platform (metadata such as episode name, audio file, and subtitles are excluded from the table for simplicity).
\begin{table*}[h]
\centering
\resizebox{\linewidth}{!}{%
\begin{tabular}{ll}
\hline
\textbf{Questions per preview} & \textbf{Response} \\ \hline
Does the preview help you decide if this episode is relevant for you? & Yes | No \\ 
Does the preview plus metadata contain enough context to understand what is being talked about from the preview? & Yes | No \\ 
Does the preview show an interesting part of the episode? & Yes | No \\ \hline
\textbf{Question per episode} & \textbf{Response} \\ \hline
Which preview is better? & Preview 1 | Preview 2 | A tie \\ \hline
\end{tabular}%
}
\caption{\label{eval_questionnaire}Offline human evaluation Label Studio questionnaire}
\end{table*}

\subsection{Online A/B Test}
The online A/B test was designed to evaluate the impact of LLM previews on user engagement and content discovery compared to legacy ML previews.

\textbf{A/B Test Context:} In the realm of digital media consumption, enhancing user engagement by facilitating content discovery is a critical focus for many platforms. We tested our LLM previews against legacy ML previews in a product that provides an interface where users can navigate through a series of podcast previews. The primary function of previews in this product is to aid users in evaluating unfamiliar podcast content, thereby enhancing podcast discovery. Improvements to these previews are expected to enhance user evaluation experience.

\textbf{A/B Test Hypothesis:} We hypothesized that LLMs would select more compelling episode previews compared to the legacy ML expert models, thus enhancing the value of each preview content and increasing the likelihood that users would have a better and more effective evaluation experience with the preview.

\textbf{A/B Test Setup:} The online A/B test was conducted over 6 weeks across 67 English-speaking countries and was available to all users in those countries. Users were evenly split between treatment and control, with users in the treatment group receiving LLM previews in the product. In order to validate the usefulness of LLM previews, we generated LLM previews for a subset of recently published English episodes, resulting in LLM previews for 34\% of episodes seen in the product during test period (remaining episodes used the same previews as control group). Users in the control group received legacy ML previews but never LLM previews. This test ran for 6 weeks, allowing sufficient time to gather meaningful data on user interactions. By implementing this test setup, we aimed to observe measurable differences in user engagement, specifically focusing on whether LLM previews could help with content discovery compared to legacy ML previews.

%% file: sections/results.tex
\section{Results}
In this section, we present the results of our offline human evaluation and online A/B test, both of which demonstrate that LLM previews outperform legacy ML previews. These findings showcase the power of LLMs in extracting more engaging and contextually rich podcast previews that improve podcast evaluation and discovery.

\subsection{Offline Human Evaluation Results}

We gathered 238 valid episode annotations to compare the performance of LLM previews against legacy ML previews.

\begin{figure}[t]
  \centering
  \includegraphics[width=\columnwidth]{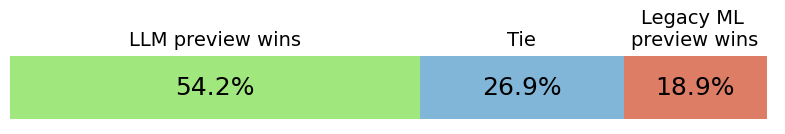}
  \caption{Offline human evaluation: Overall comparison results between LLM previews and legacy ML previews}
  \label{fig:human_eval_overall_result}
\end{figure}

\textbf{Overall Comparison Results:} The results indicated that LLM previews were better than or non-inferior to legacy ML previews 81.09\% of the time, when considering both wins and ties, and better than legacy ML previews 54.2\% of the time, when considering only wins. This implies that LLM previews were either preferred over or performed equivalently to legacy ML previews in the majority of cases (Figure \ref{fig:human_eval_overall_result}). A binomial test\footnote{Binomial test: \url{https://docs.scipy.org/doc/scipy-1.11.1/reference/generated/scipy.stats.binom_test.html}} conducted on these results yielded a p-value of 1.37e-10, allowing us to reject the null hypothesis with confidence (at a significance level of 0.001) and conclude that LLM previews' better performance is statistically significant and not due to random variation.

\begin{table}
  \centering
  \resizebox{\linewidth}{!}{
  \begin{tabular}{llll}
    \hline
    & \multirow{2}{*}{\textbf{Z-Test statistic}} & \multirow{2}{*}{\textbf{P-value}} & \textbf{LLM previews better} \\
    & & & \textbf{statistically significant?} \\
    \hline
    \textbf{Q1: Understandability} & -4.05 & 5.09e-05 & Yes \\
    \textbf{Q2: Contextual clarity} & -3.40 & 0.00067 & Yes \\
    \textbf{Q3: Interest level} & -4.32 & 1.59e-05 & Yes \\
    \hline
  \end{tabular}
  }
  \caption{\label{human_eval_question_results}
    Offline human evaluation: Question-specific results with Proportion Z-Test
  }
\end{table}

\textbf{Question-Specific Results:} Further analysis using a Proportion Z-Test\footnote{Proportion Z-Test: \url{https://www.statsmodels.org/stable/generated/statsmodels.stats.proportion.proportions_ztest.html}} on the three specific questions confirmed that LLM previews statistically significantly outperformed legacy ML previews in terms of \textit{understandability}, \textit{contextual clarity}, and \textit{interest level} (Table \ref{human_eval_question_results}).


\subsection{Online A/B Test Results}

The A/B test results indicate a marked improvement in user podcast discovery and evaluation with LLM previews over legacy ML previews. The following metrics were used to evaluate this impact:

\textbf{Podcast Evaluation Time per User:} A statistically significant improvement of 4.6\% was observed in the time users spent evaluating podcast previews during their second week in the experiment, indicating that more engaging LLM previews led to enhanced user evaluation experience.

\textbf{Evaluation Time per Preview:} LLM previews resulted in a statistically significant 4\% increase in the average time users evaluated each preview during their second week in the experiment. This improvement indicates the capability of LLMs to produce more compelling and effective preview segments, enhancing user interest and providing more value out of user evaluation period.

%% file: sections/conclusion.tex
\section{Conclusion}

In this work, we explored leveraging an LLM for podcast preview generation in a real-world application. Offline human evaluation and online A/B test results demonstrate that the LLM preview system outperforms the legacy ML system, which relies on a sophisticated aggregation of expert models. Transitioning to an LLM-based system streamlines preview generation by eliminating extensive feature engineering. This change speeds up iterations, accelerates adaptations to changing requirements, enhances scalability, and improves preview quality, highlighting the transformative power of LLMs in practical applications to simplify complex content processing tasks.



%% file: sections/ethics-statement.tex
\section{Ethics Statement}

Our system prioritizes creator autonomy through opt-out options. Podcast creators maintain full control over their content by having the option to opt out of machine-generated podcast previews (including both LLM previews and legacy ML previews). They can also generate their own previews which will replace any machine-generated previews, ensuring that their content is represented according to their preferences.

We also prioritize accessibility and inclusivity by exploring systems that can more easily adapt to diverse contexts. For instance, unlike traditional ML systems requiring extensive retraining for each language, LLMs offer better flexibility to adapt across different languages. This adaptability allows for the creation of high-quality previews that cater to diverse audiences, enhancing user experience and expanding the reach of engaging content.


%% file: sections/appendix_timestamped_transcript.tex
\section{Mock Example of Pre-Processed Transcript with Sentencization and Timestamps}
\label{sec:timestamped_transcript}

Below is a mock example showcasing the format of a pre-processed transcript with sentencization and timestamps. The transcript is first divided into individual sentences, each placed on its own line. Each sentence is accompanied by square brackets indicating the start and end timestamps in seconds. This process aids the LLM in accurately selecting previews from the episode. 

\begin{quote}
...

[01.00 - 02.50] Here is a mock sentence indicating the start of the transcript.

[03.00 - 05.25] This is another mock sentence serving as a placeholder.

[05.50 - 06.75] Yet another example of a mock sentence.

[07.00 - 09.00] This sentence is mock data for illustrative purposes.

[09.50 - 11.25] Final mock sentence to demonstrate the format.

...
\end{quote}